# High power, fixed and tunable wavelength, Grating-free Cascaded Raman fiber Lasers

V. Balaswamy[1], S. Arun[1], Santosh Aparanji, Vishal Choudhury, V.R. Supradeepa[*]

*Centre for Nano Science and Engineering, Indian Institute of Science, Bengaluru, India*
[1]*These authors contributed equally to this work*
*\*Corresponding author: supradeepa@iisc.ac.in*

**Cascaded Raman lasers enable high powers at various wavelength bands inaccessible with conventional rare-earth doped lasers. The input and output wavelengths of conventional implementations are fixed by the constituent fiber gratings necessary for cascaded Raman conversion. We demonstrate here, a simple architecture for high power, fixed and wavelength tunable, grating-free, cascaded Raman conversion between different wavelength bands. The architecture is based on the recently proposed distributed feedback Raman lasers. Here, we implement a module which converts the Ytterbium band to the eye-safe 1.5micron region. We demonstrate pump-limited output powers of over 30W in fixed and continuously wavelength tunable configurations.**

Fiber lasers have drawn significant commercial interest over the last decade due to high reliability, power scalability and robustness and are deployed widely in industrial, defense and scientific applications [1]. The superior value proposition offered by fiber lasers such as high brightness, better thermal properties, compactness and ruggedness have enabled its acceptance in these domains. However, fiber lasers based on rare-earth dopants have serious limitations. For example, high power laser emission is primarily at certain fixed wavelengths corresponding to Ytterbium (1µm), Erbium (1.5µm) and Thulium/Holmium (2µm). Also, power scalability has largely been limited to Ytterbium (Yb) emission window (1µm). Several important applications require high power fiber lasers at wavelengths other than those above, where there are no rare-earth doped fiber lasers available or limited in power scaling. For example, high power 1.5µm fiber lasers have attractive properties such as eye safety and atmospheric transparency [2]. Conventional rare-earth doped fiber lasers at 1.5µm wavelength region are based on Erbium (Er) and Erbium and Ytterbium co-doped (EY) fibers. These are limited in power scaling due to increased thermal load, progressive degradation of beam quality with power and if EY fiber is used, parasitic lasing from Yb ions when pumped at 980nm. For pumping in-band, the limitations are unavailability of low cost and high efficiency pump sources and the inability to increase gain by increasing the doping concentrations. This is due to detrimental concentration quenching and ion pair formation effects which occur at higher concentrations [3, 4]. Another serious drawback of rare-earth doped fiber lasers is their limited wavelength tuning range corresponding to the emission spectra of the constituent rare-earth ion. It would also be desirable to have widely tunable lasers at a variety of wavelength bands.

Cascaded Raman fiber lasers provide a convenient power scalable route for generating high optical powers at a variety of wavelengths [5]. This is due to the availability of gain based on stimulated Raman scattering at any arbitrary wavelength within the transmission spectrum of optical fiber. Conventionally, Raman fiber lasers used a cascaded Raman resonator (CRR) for wavelength conversion. A high power Yb doped fiber laser was used as the pump source [5]. A cascaded Raman resonator is constituted of nested cavities at each of the intermediate Stokes wavelengths and is formed with pairs of fiber Bragg gratings with a high nonlinearity fiber referred to as Raman fiber in between. Even though conventional Raman fiber lasers are simple and straight forward to implement, they are limited in power scaling due to the inherently lossy nature of CRR. More than 100W at 1480nm has been demonstrated with an optical to optical conversion efficiency of ~ 48% starting from 1117nm [6], reasonably lower than the quantum limited efficiency of 75%. Further, these lasers tend to become unstable at higher power operations showing oscillations in the output power due to coupling between the cascaded Raman resonator and the rare-earth doped fiber laser [7]. In 2013, an efficient and reliable architecture was proposed which works based on the single pass cascaded Raman amplification [8] seeded at all the intermediate Stokes wavelengths with a low power cascaded Raman laser. An output power of more than 300W was demonstrated at a conversion efficiency from 1117nm to 1480nm of 64% [9]. However, the system required an additional secondary, lower power cascaded Raman laser to seed the intermediate Stokes wavelengths to achieve conversion in a single pass. This made the system more complex. This issue was overcome recently with a simplified, all passive cascaded Raman conversion module. Here, a fraction of the power from the main laser is used to power the secondary cascaded Raman laser necessary for generating the intermediate Stokes wavelengths [10].

However, in all the previous work on cascaded Raman lasers, their input and output wavelengths are fixed and decided by the fiber Bragg gratings used in the cascaded Raman conversion module. This is a rather serious limitation since a new module is necessary when the input wavelength is changed. Such an approach is not practical since this requires different fiber Bragg grating sets for each input-output combination. Fabrication process of such high power handling cascaded grating sets is complicated and hence such an approach is not convenient and cost effective. It is highly desirable to have a passive, cascaded Raman conversion module which is completely color blind to the input pump source, and should generate the desired output wavelength with high power and reliability. Recently, such a passive cascaded Raman conversion module, independent of input wavelength has been proposed by Babin et al [11, 12]. They have demonstrated a cascaded Raman fiber laser, based on distributed random feedback due to Rayleigh back scattering occurring all along the length of the passive Raman gain fiber. Since the distributed feedback due to Rayleigh back scattering is present at any arbitrary wavelength, such systems naturally provide wavelength flexibility as their output wavelength can be changed by just changing the input pump wavelength. These systems have been further developed by the group of Yan Feng [13, 14]. More details on random distributed feedback fiber lasers is available in [15].

However, the above mentioned distributed feedback cascaded Raman fiber lasers are limited in power scaling. The reason for limited power scaling is either due to conversion of the desired wavelength to the next higher order stokes and/or due to the generation of a supercontinuum as the cascaded stokes conversion moves to the anomalous dispersion region of the passive fiber used for Raman gain. For example, a maximum of ~10W in a 3 rd order Raman laser using a polarization-maintaining
fiber at 1.23μm wavelength in [12] and <8W across the bandwidth in a Raman laser [14] has been demonstrated. Thus, in contrast to output powers achieved with other types of cascaded Raman lasers previously, the output powers with distributed feedback based Raman lasers is limited. In this work, we overcome these problems and demonstrate a high power, grating-free, cascaded Raman conversion module which is independent of the input pump wavelength. This proposed module provides a technique to perform conversion from an arbitrary input wavelength band into a longer arbitrary output wavelength band provided it is in the transmission region of the optical fiber. Here, we implement a module which converts the Ytterbium band (1050-1120nm) to the eye-safe 1.5micron region. We demonstrate pump-limited output powers of over 30W in fixed and continuously wavelength tunable configurations with the same module.

Figure 1 shows the schematic of the module. The input consists of a high power Yb doped fiber laser. Any conventional Ytterbium doped fiber laser off the shelf can be utilized for this purpose. In this work, we use three different sources. Two fixed wavelength Ytterbium lasers operating at 1117nm and 1080nm and a tunable laser operating in the emission band of Ytterbium. A long length (~350m) of high Raman nonlinearity fiber with additional filtering properties is used. This fiber, referred to as Raman filter fiber [6] is used to provide Raman gain as well as random distributed feedback. The unique properties of the Raman filter fiber, as shall be shown enable the overcoming of issues faced by previous work. The output end of the Raman filter fiber is cleaved at an angle >$8^0$ to avoid Fresnel backward reflection. A fused fiber wavelength division multiplexer (WDM) is used to launch the pump source into the fiber. Distributed backscatter in the shifted wavelengths is also efficiently separated by the WDM and recoupled back to enable seeded, preferential forward cascaded Raman conversion. Since Raman gain is large, a simple flat-cleave which provides a 4% Fresnel reflection is sufficient for feedback as shown in the figure. The Flat cleave along with Raman filter fiber forms a half – open random cavity and enables high efficiency cascaded Raman conversion. The WDM used is a 1117/1480nm WDM which works between the input band at Ytterbium wavelengths and the 1.5micron region. The choice of the WDM is not exact and any WDM operating between the two bands can be used.

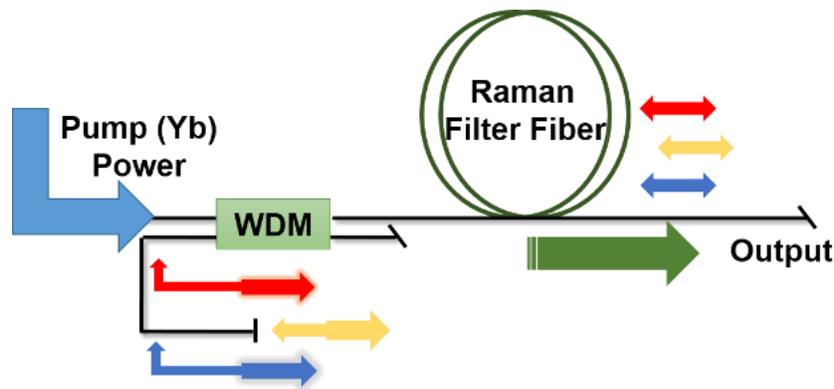

Fig. 1. Schematic of the grating – free, passive, cascaded Raman conversion module; WDM – wavelength division multiplexer.

The Raman filter fiber used eliminates the two issues mentioned previously which hampered power scaling. The fiber has a cut-off wavelength beyond which the transmission loss in the fiber becomes very high (> 10dB/m). This effectively terminates the cascade to the required wavelength band adjacent to the cut-off wavelength. The filter fiber is based on a W-shaped index profile which when coiled provides an enhanced wavelength dependent loss profile due to bending. The fiber used in our experiment is from OFS and was coiled to a 9cm diameter which resulted in high distributed bending losses for wavelengths greater than 1520nm [8]. Further, the Raman filter fiber used is of low effective area (12μm$^2$), and high NA (~0.2) due to which, the zero dispersion wavelength is shifted towards longer wavelengths (>1800nm) and provides very high normal dispersion of ~-80ps/nm.km over a broad wavelength range. This property prohibits the generation of supercontinuum by not allowing nonlinearities such as modulational instability which is needed to kickstart this process.

Fig. 2 shows the results of the first experiment when the grating-free module is pumped with a fixed wavelength laser at 1117nm. The output power Fig 2(a) and spectral characteristics Fig 2(b) are shown. Since the cutoff of the filter fiber is at 1520nm, this source converts 1117nm to 1480nm corresponding to a 5$^{th}$ order cascaded Raman convertor. The output power characteristics exhibit a typical laser behavior with almost zero output 1480nm power till the threshold is reached. After the threshold, the output 1480nm power grows quickly with the input pump power. The next higher order stokes of 1480nm is 1590nm which falls in the high loss region of the Raman filter fiber. Here, the Raman gain is far below the loss and thereby its generation is suppressed. Fig 2(b) shows the normalized output spectrum at full power in linear scale. More than 95% of output power is present in the final stokes indicating a very high degree of wavelength conversion. The maximum input pump power at 1117nm coupled into the high power, grating–free cascaded Raman converter was ~90W, for which a maximum output power of ~43W at 1480nm was obtained. This corresponds to a conversion efficiency of ~48% from input 1117nm pump to output 1480nm laser signal.

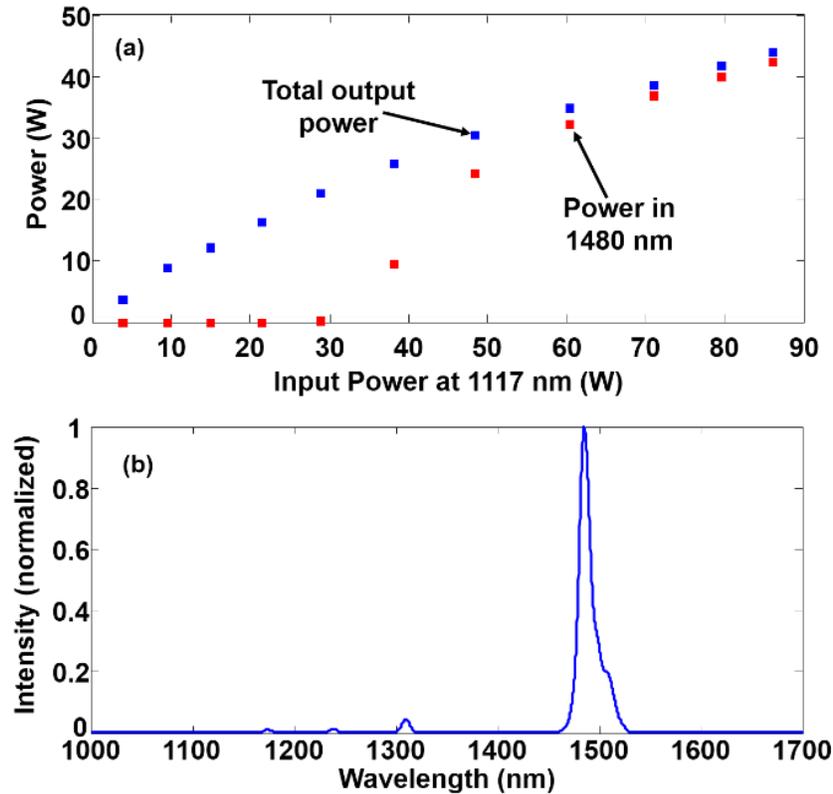

Fig. 2. Results with pumping at a fixed wavelength of 1117nm (a) Output power vs Input coupled power (b) spectrum in linear scale.

The WDM used in our experiments was non-ideal and provides more than 1dB of insertion and cross coupling loss to the input pump signal thereby reducing the coupled pump power into the Raman conversion module. Also, the splice between the WDM and the Raman filter fiber was not optimized which also provides considerable loss to the pump signal. The efficiency can be further enhanced by addressing these issues. The 4% Fresnel reflection at the flat cleave is sufficient enough to seed the cascaded conversion entirely in the forward direction and the backward power coming from the flat cleave is negligible compared to forward generated power. This has already been demonstrated in [16].

In order to demonstrate the agility of our cascaded Raman conversion module, we have used another high power Yb doped fiber laser operating at 1080nm as a pump source. This resulted in a 6$^{th}$ order cascaded Raman conversion to 1511nm. Fig. 3(a) shows the normalized spectrum for two input pump sources used at 1080nm (blue) and 1117nm (red). Fig. 3(b) shows the output wavelength band around 1.5micron. Wavelength conversion to the final stokes of 1480nm and 1511nm has been achieved. Fig. 3(c) shows the plot of normalized pump and Raman spectra together indicating the wavelength band conversion from 1μm to 1.5μm. The conversion efficiency achieved from 1080nm to 1511nm was ~41% with an output power of ~33W for input pump power of 85W. The output power of the 1080nm source was lower than the 1117nm source resulting in reduced coupled power. The corresponding quantum limited conversion efficiency from 1080nm to 1511nm is ~71%. The demonstrated output powers are currently limited only by the available input pump power.

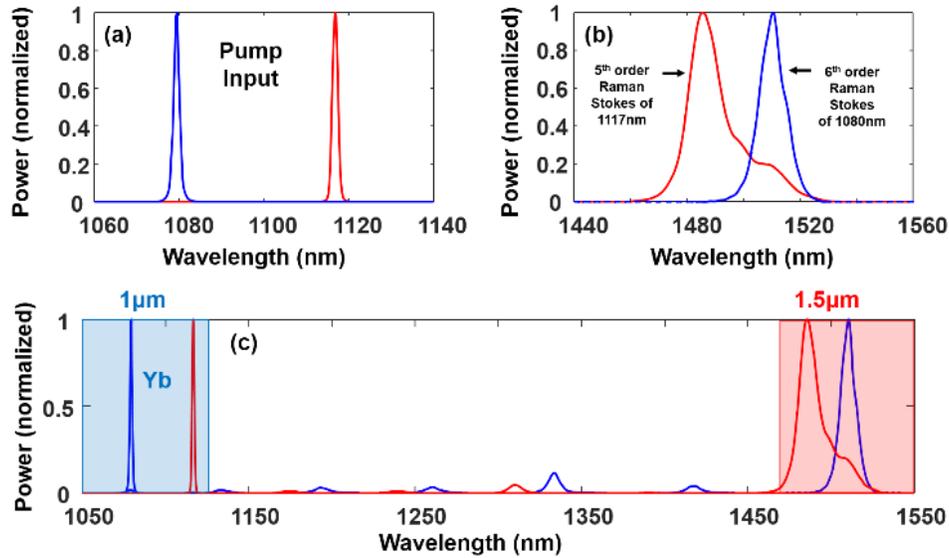

Fig. 3. (a) Normalized Spectrum showing input (pump) wavelengths at 1080nm (blue) and 1117nm (red) and (b) Output spectrum corresponding to 1080nm (blue) and 1117nm (red) pump. (c) Both pump and Raman spectra shown together indicating the band conversion from 1μm to 1.5μm wavelength.

Recently, using a similar, cascaded Raman conversion module based on distributed feedback, an output power of 100.1W at 1806nm has been demonstrated [17]. This corresponds to 9th order Raman stokes starting from a fixed pump wavelength of 1064nm. Here, the authors utilize conventional Raman fiber and have exploited the enhanced inherent losses of Silica Raman fiber at longer wavelengths as a natural filter fiber. However, with this principle, the cut-off wavelength and thus optimal operating band is fixed. In this work, by the choice of a designed filter fiber, an arbitrary cut-off wavelength and operating band can be achieved.

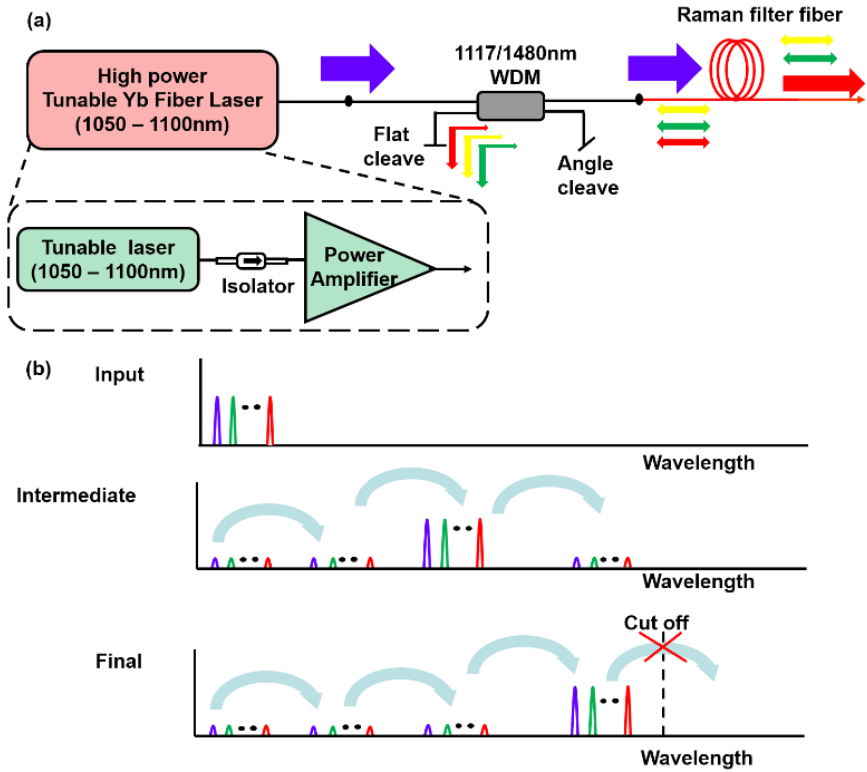

Fig. 4. (a) Schematic of high power, tunable, grating-free cascaded Raman fiber laser. (b) Schematic representation of mechanism of tunable grating-free Raman fiber laser.

In order to demonstrate wavelength tunability with the same cascaded Raman conversion module, we have used an in-house built high power, continuously tunable (1050-1100nm) Yb doped fiber laser [18] as a pump source. Fig. 4(a) shows the schematic of the architecture, a tunable Yb fiber laser based on a master oscillator power amplifier (MOPA) configuration (shown in inset) is the pump source. In the current experiment, the pump source could consistently couple between 85 to 90W into the cascaded Raman conversion module at all its operating wavelengths. Fig 4(b) shows the operating principle of tunable cascaded Raman fiber laser. The tunable Yb doped fiber laser undergoes Raman cascading to longer and longer wavelengths, till it gets terminated by the cut-off of the filter fiber. This results in the band conversion of the input Ytterbium band to the output band decided by the cut-off of the filter fiber. In our experiment, we operated the tunable laser from 1065-1100nm in steps of 5nm to full power and the corresponding output spectra was recorded. In our experiment, because the cut-off of the Raman filter fiber at 1520nm which corresponds to an input pump wavelength of 1085nm, 6th order Raman stokes shift is observed for input pump wavelengths till 1085nm and 5th order Raman stokes shift is observed for pump wavelengths greater than 1085nm.

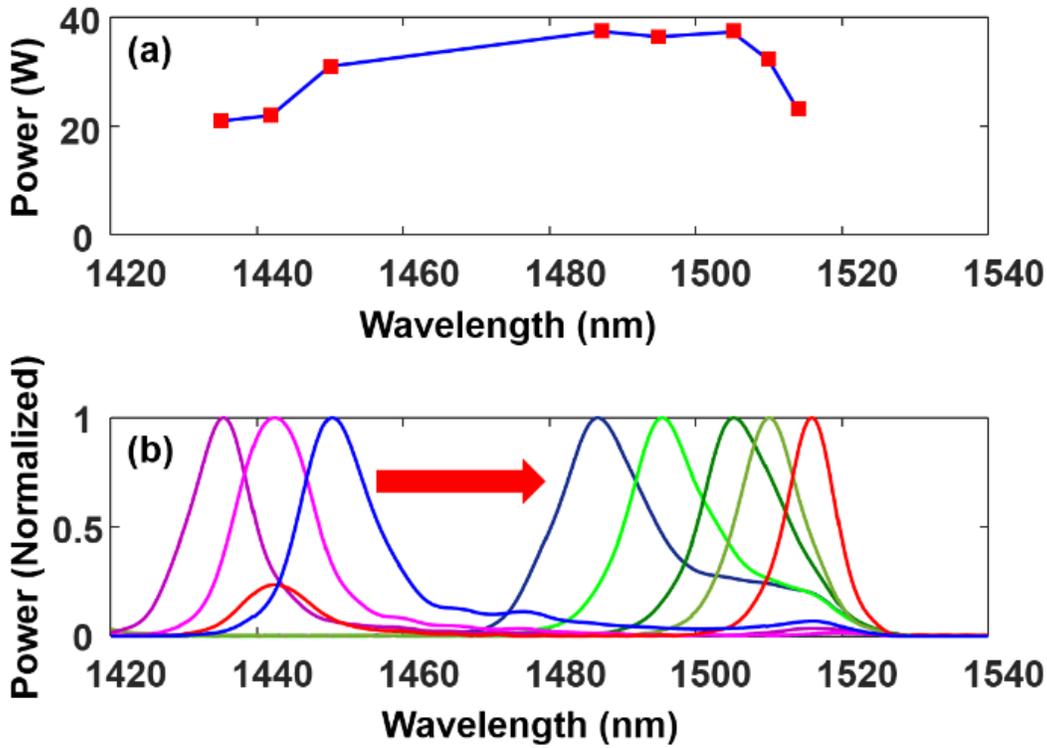

Fig. 5. (a) output power vs wavelength in the tunable, grating-free cascaded Raman laser. (b) Normalized output spectra vs wavelength.

Fig. 5 shows the wavelength tuning characteristics. A tuning range from 1450-1510nm with >30W of in-band power was measured (Fig. 5(a)). Even though our pump wavelength tuning range (1050-1100nm) corresponds to one Raman shift, in experiments, we have operated it from 1065-1100nm. This resulted in a gap between 1450nm to 1486nm in output tuning range. For pump wavelengths between 1050-1065nm, we did not attempt full power scaling due to the heating of the isolator from enhanced ASE noise. By suitable choice of the isolator and improved amplifier design, the gaps in the tuning of output wavelength can be eliminated. The output powers are limited by the input pump power. To the best of our knowledge, achieved output power values have been the highest in tunable cascaded Raman fiber lasers. Fig. 5(b) shows the normalized output spectrum in linear scale. The broadening of the spectrum is wavelength specific and it is seen that at the middle, the broadening is maximal. The reason is that at these wavelengths, the output power is comparatively higher and the cut-off of the filter fiber is farther away from the emission wavelength. Since the Raman gain is substantial over a wide bandwidth, this results in a characteristic Raman gain broadened shape. As the wavelength get closer to the filter fiber cut-off or the power reduces, the linewidth narrows reasonably.

In summary, we have demonstrated an all passive, grating-free, cascaded Raman conversion module which can convert any Yb doped fiber laser operating at 1μm wavelength region to 1.5μm wavelength region. We demonstrated the module with fixed wavelength, high power, grating-free, cascaded Raman fiber lasers at 1480nm and 1511nm with 43W and 33W of output power by using fixed wavelength Yb fiber lasers at 1117nm and 1080nm. And by using a high power tunable Yb fiber laser as pump source, high power tunable cascaded Raman fiber laser with >30W of output power between 1450-1510nm was demonstrated. In this work we have demonstrated the wavelength band conversion from 1μm wavelength region to 1.5μm wavelength region, however, this approach can be easily modified (through the modification of the cut-off wavelength of the filter fiber) to achieve high power wavelength band conversion at any arbitrary wavelength region within the transmission window of the silica fiber.

**Funding.** Department of Science and Technology, Government of India, (SB/S3/EECE/0149/2015, INSPIRE - IFA14-ENG-78)

**Acknowledgements.** Jeffrey W. Nicholson (OFS Laboratories).

## References

1. D. J. Richardson, J. Nilsson, and W. A. Clarkson J. Opt. Soc. Am. B 27, B63- B92 (2010).
2. https://lia.org/publications/ansi/Z136–1
3. V. Kuhn, D. Kracht, J. Neumann, and P. Weßels, Opt. Lett. 36, 3030 (2011).
4. J. Zhang, V. Fromzel, and M. Dubinskii, Opt. Express 19, 5574 (2011).
5. V. R. Supradeepa, Y. Feng, J. W. Nicholson, Journal of Optics, 19, (2017).
6. V. R. Supradeepa, J. W. Nicholson, C. E. Headley, Y. Lee, B. Palsdottir, D. Jakobsen, no. 8237-48, SPIE photonics west 2012.
7. J. W. Nicholson, M. F. Yan, P. Wisk, J. Fleming, F. DiMarcello, E. Monberg, T. Taunay, C. Headley, and D. J. DiGiovanni, Opt. Lett. 35(18), 3069–3071 (2010).
8. V. R. Supradeepa, Jeffrey W. Nichsolson, Clifford E. Headley, Man F. Yan, Bera Palsdottir, and Dan Jakobsen, Opt. Express 21, 7148-7155 (2013).
9. V. R. Supradeepa and Jeffrey W. Nicholson. Opt Lett 38, 2538- 2541 (2013).
10. V. R. R. Supradeepa, V. Balaswamy, S. Arun, and G. Chayran, in Conference on Lasers and Electro-Optics, OSA Technical Digest (2016) (Optical Society of America, 2016), paper SM2Q.5.
11. Babin, S. A., Vatnik, I. D., Laptev, A. Y., Bubnov, M. M. & Dianov, E. M. Opt. Express 22, 24929 (2014).
12. S. A. Babin, E. A. Zlobina, S. I. Kablukov, and E. V. Podivilov, Sci. Rep. 6, 22625 (2016).
13. Lei Zhang, Huawei Jiang, Xuezong Yang, Weiwei Pan, and Yan Feng, Opt. Lett. 41, 215-218 (2016).
14. Lei Zhang, Huawei Jiang, Xuezong Yang, Weiwei Pan, Shuzhen Cui & Yan Feng, Sci. Rep. 7, 42611(2017).
15. S. K. Turitsyn, S. A. Babin, D. V. Churkin, I. D. Vatnik, M. Nikulin, and E. V. Podivilov, Phys. Rep. 542, 133–193 (2014).
16. H. Wu, Z. Wang, M. Fan, L. Zhang, W. Zhang, and Y. Rao, Opt. Express 23(2), 1421–1427 (2015).
17. Zhang, L., Dong, J. and Feng, Y., IEEE J. Sel. Top. Quantum Electron. 24(3) (2018).
18. V. Balaswamy, S. Aparanji, G. Chayran, and V. R. R. Supradeepa, in 13th International Conference on Fiber Optics and Photonics, OSA Technical Digest (online) (Optical Society of America, 2016), paper Tu3E.4